\documentclass[fleqn,usenatbib]{mnras}
\usepackage{epsfig}
\usepackage{amsfonts}
\usepackage[usenames]{color}
\usepackage[usenames,dvipsnames]{xcolor}
\usepackage{url}
\usepackage{subfigure}
\usepackage{times,graphicx,amsmath,amsfonts,amssymb,aas_macros,epstopdf,hyperref}
\usepackage[normalem]{ulem}
\usepackage[T1]{fontenc}
\usepackage{multirow}
\usepackage{aecompl}
\usepackage{newtxtext,newtxmath}
\usepackage{bm,upgreek}
\usepackage{booktabs, blkarray}

\DeclareRobustCommand{\VAN}[3]{#2}
\let\VANthebibliography\thebibliography
\def\thebibliography{\DeclareRobustCommand{\VAN}[3]{##3}\VANthebibliography}

\newcommand{\yt}{\textcolor{black}}

\usepackage{hyperref}
\hypersetup{
    colorlinks=true,
    linkcolor=blue,
    citecolor=blue,
}

\def\etal{{\frenchspacing\it et al.}}
\def\ie{{\frenchspacing\it i.e.}}
\def\eg{{\frenchspacing\it e.g.~}}
\def\etc{{\frenchspacing\it etc.}}
\def\d{{\rm d}}

\def\be{\begin{equation}}
\def\ee{\end{equation}}
\def\ba{\begin{eqnarray}}
\def\ea{\end{eqnarray}}

\def\LaTeX{L\kern-.36em\raise.3ex\hbox{a}\kern-.15em
    T\kern-.1667em\lower.7ex\hbox{E}\kern-.125emX}

\title[BAO and RSD in anisotropic CF]{Extracting key information from spectroscopic galaxy surveys}
\author[Wang \etal]{
\parbox{\textwidth}{
Yuting Wang$^{1,2}$\thanks{\url{Email: ytwang@nao.cas.cn}}, Gong-Bo Zhao$^{1,3,2}$\thanks{\url{Email: gbzhao@nao.cas.cn}}, John A. Peacock$^{4}$\thanks{\url{Email: jap@roe.ac.uk}}} 
\vspace*{30pt} \\
$^{1}$ National Astronomy Observatories, Chinese Academy of Science, Beijing, 100101, P.R.China \\
$^{2}$ Institute for Frontiers in Astronomy and Astrophysics, Beijing Normal University, Beijing 102206, P.R.China \\
$^{3}$ School of Astronomy and Space Sciences, University of Chinese Academy of Sciences, Beijing 100049, P.R.China \\
$^{4}$ Institute for Astronomy, University of Edinburgh, Royal Observatory, Edinburgh EH9 3HJ, United Kingdom
}

\date{Accepted XXX. Received YYY; in original form \today}

\pubyear{2024}

\begin{document}
\label{firstpage}
\pagerange{\pageref{firstpage}--\pageref{lastpage}}
\maketitle

\begin{abstract} 
We develop a novel method to extract key cosmological information, which is primarily carried by the baryon acoustic oscillations and redshift space distortions, from spectroscopic galaxy surveys based on a joint principal component analysis (PCA) and massive optimized parameter estimation and data compression (MOPED) algorithm. We apply this method to galaxy samples from BOSS DR12, and find that a PCA manipulation is effective at extracting the informative modes in the 2D correlation function $\xi(s, \mu)$, giving a tighter constraint on BAO and RSD parameters compared to that using the lowest three multipole moments by the traditional method; \ie\, the Figure of Merit of BAO and RSD parameters is improved by \yt{$17\%$}. We then perform a compression of the informative PC modes for BAO and RSD parameters using the MOPED scheme, reducing the dimension of the data vector to the number of interesting parameters, manifesting the joint PCA and MOPED as a powerful tool for clustering analysis with almost no loss of constraining power.
\end{abstract}


\section{Introduction}
\label{sec:intro}
The effects of baryon acoustic oscillations (BAO) \citep{Eisenstein2005,Cole:2005sx,Percival2006} and redshift space distortions (RSD) \citep{Kaiser1987, Peacock:2001gs}, which shape specific three-dimensional clustering patterns of galaxies, can be used for probing the background expansion and the structure growth of the universe respectively; these are crucial for cosmological studies, including tests of dark energy or gravity theories \citep{Weinberg2013, Koyama:2015vza}. This makes measurements of BAO and RSD one of the key science drivers of massive spectroscopic galaxy surveys, for example, the Sloan Digital Sky Survey (SDSS) \citep{SDSS}, the Two-Degree-Field Galaxy Redshift Survey (2dFGRS) \citep{2dFGRS}, WiggleZ \citep{WiggleZ}, the SDSS-III Baryon Oscillation Spectroscopic Survey (BOSS) \citep{Dawson2013}, and the SDSS-IV extended Baryon Oscillation Spectroscopic Survey (eBOSS) \citep{Dawson2016}.

The two-point clustering of galaxies, which contains primary information of BAO and RSD, is generically quantified by either the anisotropic correlation function (CF) $\xi(s, \mu)$, or power spectrum (PS) $P(k, \mu)$, where $s$ and $k$ denote the separation of pairs in the configuration space or Fourier space, respectively, and $\mu$ is the cosine of the angle between the line of sight (LOS) and either the separation vector of two tracers or the wave vector. CF or PS can be represented by a combination of multipoles of various orders, and it was found that using the monopole and quadrupole is almost sufficient for a measurement of the anisotropic BAO \citep{Ross:2015mga}. For a joint BAO and RSD measurement, it is a common practice to include the hexadecapole in the analysis, which is the highest non-vanishing multipole on linear scales \citep{Kaiser1987}. Higher-order moments are generally informative and complementary to the lower-order moments. An ideal analysis will maximally extract the cosmological information from all the moments with controlled systematics and computational costs. For this purpose, it is crucial to separate the signal-dominated modes from the noise-dominated ones.

One approach is to perform a principal component analysis (PCA) \citep{pca1, pca2} on the measured two-point statistics. As we shall discuss, a PCA manipulation can efficiently extract the informative modes from the measured $\xi(s, \mu)$, and yield a tighter constraint on BAO and RSD parameters compared to that using the low-order multipole moments, \ie\, $\xi_{\ell}(s)$ with moments of $\ell \leq 4$.

As we shall demonstrate in this work, in configuration space the squashing effect along the LOS caused by RSD in the 2D correlation function, $\xi(s, \mu)$, manifests a non-local feature, which can be easily recovered by the first few principal modes. In contrast, the BAO signal manifests itself as a local bump on scale of $\sim 150\,\rm Mpc$ in 2D correlation function: thus more principal modes are required in order to reconstruct this local feature. Fortunately, the massive optimized parameter estimation and data compression (MOPED) algorithm \citep{MOPED} can be used to perform a compression to reduce the number of principal modes with almost no loss of constraining power. \yt{The relevant data compression has been successfully applied into the analysis of galaxy clusterings in Fourier space \citep{Gualdi:2017iey,Lai:2023nil}.}

In this work, we develop an efficient method to extract key cosmological information for BAO and RSD from galaxy surveys based on a joint PCA and MOPED scheme. We apply our method to the isotropic and anisotropic correlation functions. The analyzing results prove that a joint PCA and MOPED manipulation is effective at extracting BAO and RSD in the 2D correlation function.

The paper is structured as follows: the details on the method are described in Sec. \ref{sec:method}. In Sec. \ref{sec:dem}, we demonstrate our methodology using the mock and galaxy samples of the BOSS data release (DR) 12, and present results of applying our method. We conclude in Sec. \ref{sec:con}.

\section{Methodology} 
\label{sec:method}
\subsection{Principal components} 
Let ${\bf C}$ be the data covariance of the two-point CF of the galaxies, with entries 
\ba \label{eq:cov}
\yt{C_{qq'}\equiv \frac{1}{N_{\rm mock}-1} \left\langle \left[\xi(S_q)-\bar{\xi}(S_q)\right]\left[\xi(S_{q'})-\bar{\xi}(S_{q'}) \right] \right\rangle \,,}
\ea 
where, \yt{$S_q$ is the separation of galaxy pairs, the indices denote the discretized separation bins, and} the overbars denote the mean value of CF, defined as
\ba \label{eq:mean}
\yt{\bar{\xi}(S_q)  = \frac{1}{N_{\rm mock}} \sum_{n=1}^{N_{\rm mock}} \xi_n(S_q)\,,}
\ea
and $N_{\rm mock}$ is the number of mocks used. The expressions (Eqs.\,\ref{eq:cov} and \ref{eq:mean}) are general for both the isotropic CF (\ie, the monopole of CF), where \yt{$\xi({S})=\xi_0(s), \ s \equiv |S|$}, and for the anisotropic CF, \yt{where $\smash{\xi(S)=\xi(s,\mu)}, \mu$ is the cosine of the angle between the separation vector and the line of sight.} For the former case, the indices denote the discretized $\xi$ in bins of $s$, while for the latter, the indices mark the pixillized $\xi$ in $s$ and $\mu$. 

We then diagonalize ${\bf C}$, namely, 
\ba\label{eq:PCA} 
{\bf C}={\bf W}^T {\bm \Lambda} {\bf W}, 
\ea 
where the diagonal matrix ${\bm \Lambda}$ and the decomposition matrix ${\bf W}$ store the eigenvalues $\lambda_i$ and the orthonormal eigenvectors $e_i(S)$ respectively. The observed data vector $\xi(S)$ can then be expanded in the orthonormal basis $\{e_i(S)\}$, with the $i$th expansion coefficient and its uncertainty being 
\ba \label{eq:coeff} 
\yt{\gamma_i = \sum_{q=1}^{N_{S}} \xi(S_q) \cdot e_i(S_q),}
\ea
and $\smash{\sigma(\gamma_i)=\sqrt{\lambda_i}}$, respectively. Here, \yt{$N_{S}$ is the number of the separation bins. Specifically,} for the monopole of CF, \yt{$N_{S} \equiv N_{s}$} is the number of $s$ bins. And for the anisotropic CF,  \yt{\smash{$N_{S} \equiv N_{s} \times N_{\mu}$}} is the number of $(s, \mu)$ pixels. Then we demonstrate that \yt{most of} the information in $\xi(S)$, can be extracted using a small fraction of the $\gamma$'s with the largest variances, which are uncorrelated variables by construction, as observables.

Generally, the eigenvectors can be used to reconstruct various kinds of observables, including the original anisotropic CF, namely, 
\ba\label{eq:xismu} 
\xi(s,\mu) = \sum_{i=1}^{M} \gamma_i \ e_i(s,\mu), 
\ea 
and the widely-used CF multipoles and wedges, 
\ba\label{eq:xiIs} 
\xi_{\rm I}(s) =  \sum_{i=1}^M \gamma_i \ e_{i,{\rm I}}(s), 
\ea 
where, \yt{$M$ is the number of the eigen modes, and} 
\ba\label{eq:eiIs} 
e_{i,{\rm I}}(s) = \frac{1}{N_{\rm I}} \int_{\mu_{\rm min}}^{\mu_{\rm max}} {\rm d}\mu \  e_i(s,\mu) Q_{\rm I}(\mu). 
\ea 
Note that for the CF multipoles, $Q_{\rm I}(\mu)$ are the Legendre polynomials, $P_{\ell}(\mu)$, $\smash{N_{\rm I}=2/(2\ell+1)}$ and $\smash{\mu_{\rm min}=-1,\,\mu_{\rm max}=1}$. For the wedges, $\smash{Q_{\rm I}(\mu)=1}$, $\smash{N_{\rm I}=\mu_{\rm max}-\mu_{\rm min}}$. The covariance between the discretized $s$ bins can then be evaluated as, \ba\label{eq:covxi} \yt{{\rm cov}\left[\xi_{\rm I}(s_q),\xi_{\rm I'}(s_{q'})\right] =\frac{1}{N_{\rm I}N_{\rm I'}} \sum_{i=1}^M \lambda_i  \ e_{i,{\rm I}}(s_q) e_{i,{\rm I'}}(s_{q'}).}\ea From Eqs. (\ref{eq:xismu}) - (\ref{eq:covxi}), we can see that the principal component (PC) modes $e_i$ form a natural reservoir to store the information for $\xi(s,\mu)$, which enables a rapid comparison of results derived from various kinds of observables. 

\subsection{Fisher information matrix} 
Based on the PC modes, the Fisher information matrix for the parameters $\theta_a$ and $\theta_b$ using CF as observables can be written as, 
\ba 
F_{ab} =\sum_{i=1}^{M} \frac{(\gamma_{i})_{,a} (\gamma_{i})_{,b}}{\lambda_i } \approx \sum_{i=1}^{M'} \frac{(\gamma_{i})_{,a} (\gamma_{i})_{,b}}{\lambda_i },
\ea 
where $M' < M$ denotes the first $M'$ eigen modes. As we shall demonstrate later, using a small fraction of the PC modes is sufficient to extract \yt{most of the information about $\theta_a$ in two-point CF}. $(\gamma_{i})_{,a}$ is the derivative of the $i$th expansion coefficient with respect to the parameter $\theta_a$, \ie\,
\ba
(\gamma_{i})_{,a}=\frac{\partial \gamma^{\rm th}_{i} }{\partial \theta_a}\,,
\ea
where $\gamma_i^{\rm th}$ denotes the $i$th expansion coefficient of the theoretical CF, \ie, $\smash{\gamma_i^{\rm th}=\xi^{\rm th}(S)\cdot e_i(S)}$, as the dependence on the parameters is computed by the theoretical model of CF.  In Sec. \ref{sec:dem}, we describe the models for isotropic CF and anisotropic CF we use in detail.

The Fisher matrices using multipoles or wedges can be obtained similarly, and are simply algebraic expressions including $(\gamma_{i})_{,a}, (\gamma_{i})_{,b},\lambda_i$, and $e_{i,{\rm I}}$. This makes it straightforward to derive parameter constraints from the multipoles or wedges without starting from scratch. 

\begin{figure} %[!t]
\centering
\includegraphics[scale=0.21]{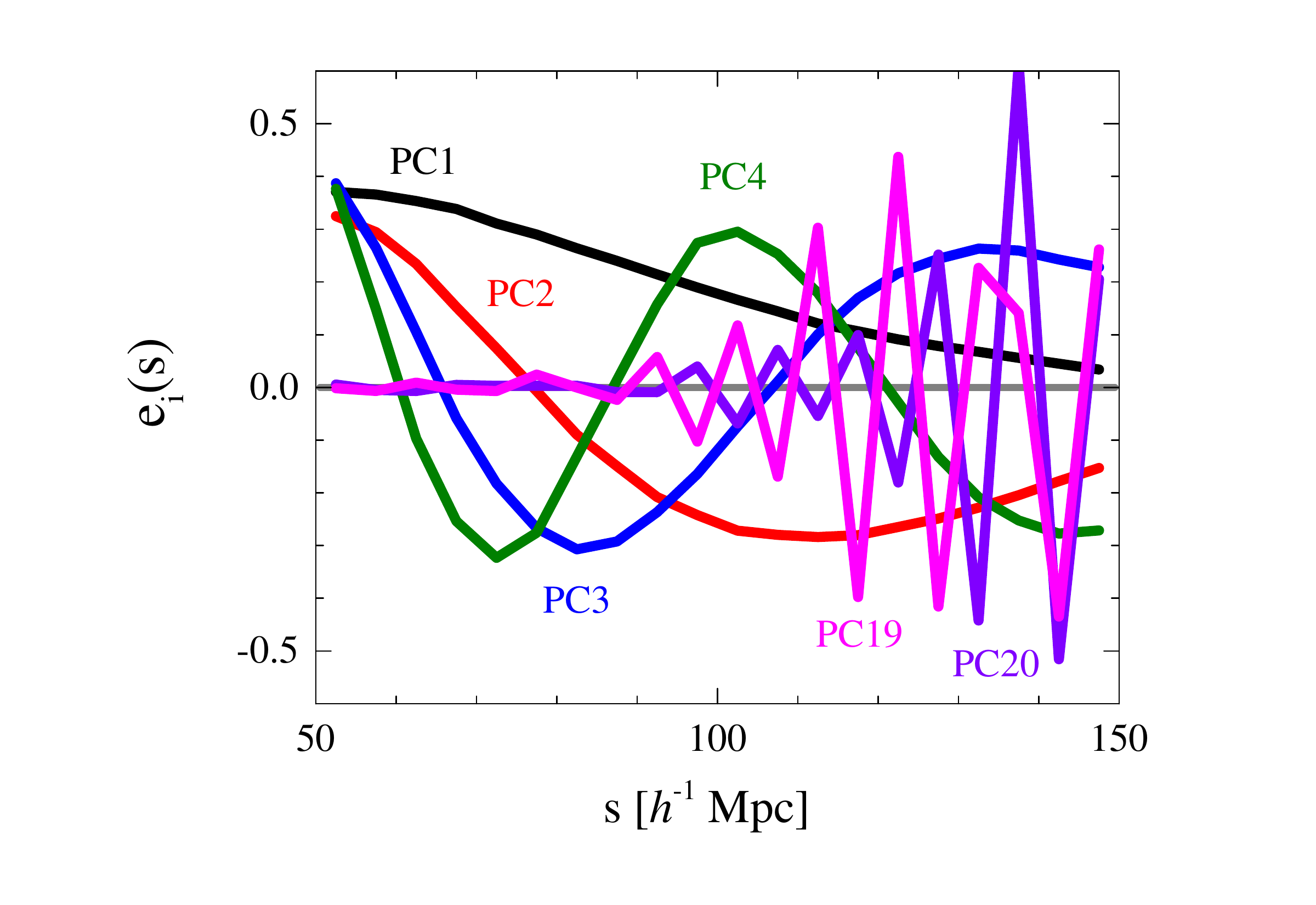}
\caption{\yt{The first four (with largest variances) and the last two (with least variances) eigenvectors of the data covariance matrix for the monopole of the galaxy correlation function.}}
\label{fig:fig1}
\end{figure} 

\begin{figure} %[!t]
\centering
\includegraphics[scale=0.21]{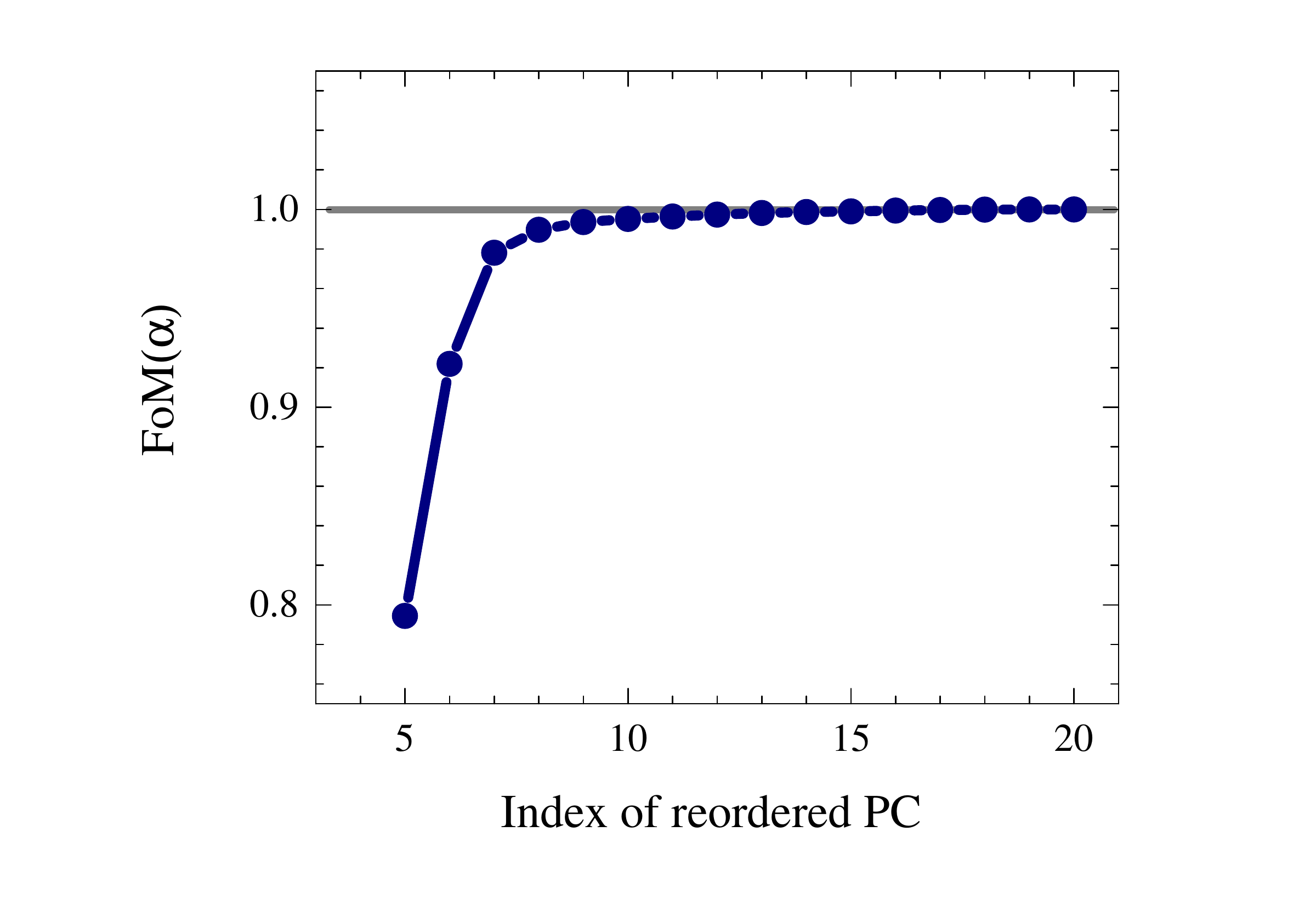}
\caption{\yt{The forecasted FoM of the isotropic BAO parameter $\alpha$, as a function of the number of total reordered PC modes used. The FoM is normalized so that it is exactly $1$ if all the PC modes are used.}}
\label{fig:fig2}
\end{figure} 

\begin{figure} %[!t]
\centering
\includegraphics[scale=0.21]{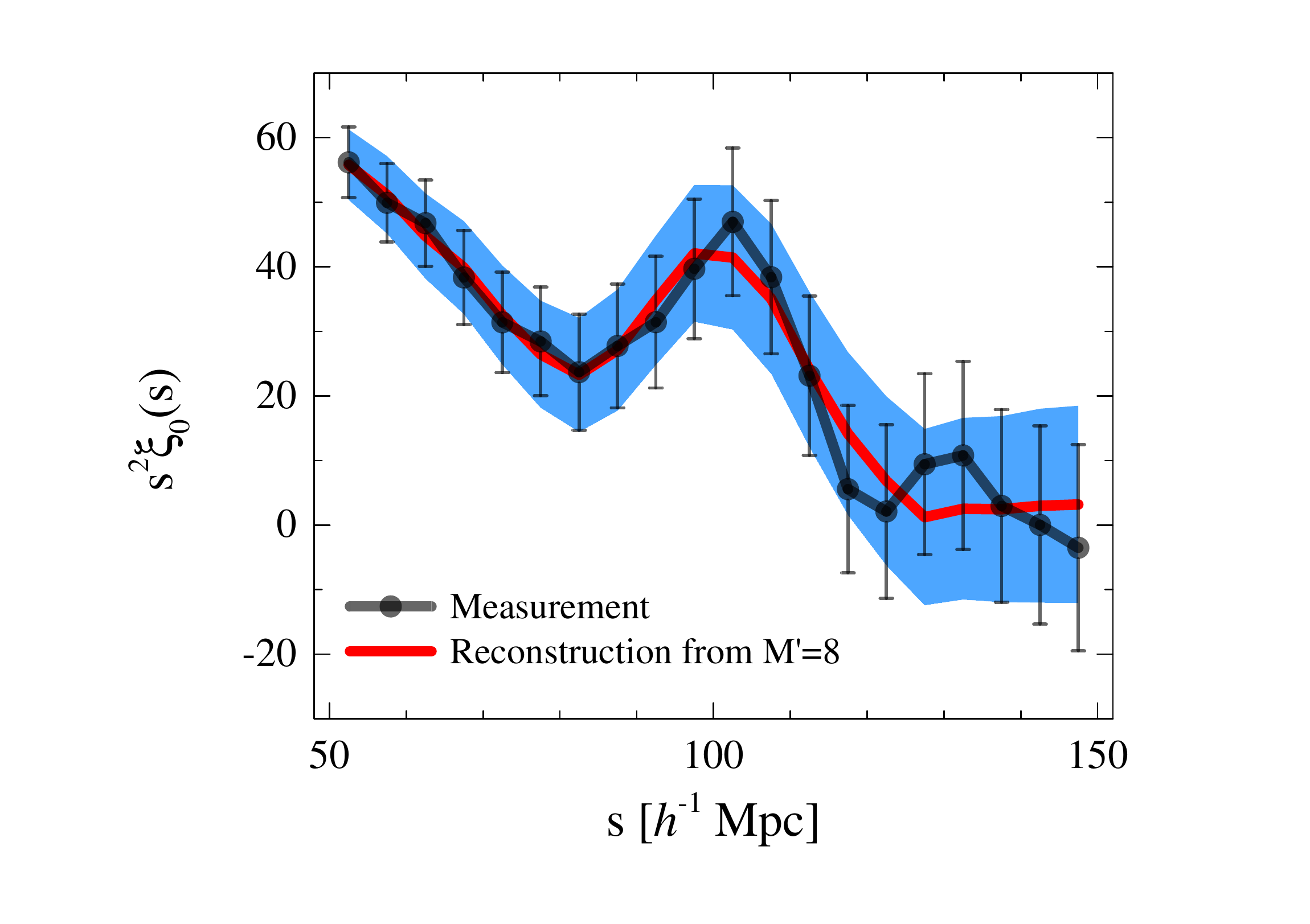}
\caption{\yt{A comparison between the measured $\xi_0$ (black curves with the 68\% CL error bars) and the reconstructed $\xi_0$ from the first eight ($M'=8$) PC modes (red curve with a blue band showing the 68\% CL uncertainty).}}
\label{fig:fig3}
\end{figure} 

\begin{figure} %[!t]
\centering
\includegraphics[scale=0.21]{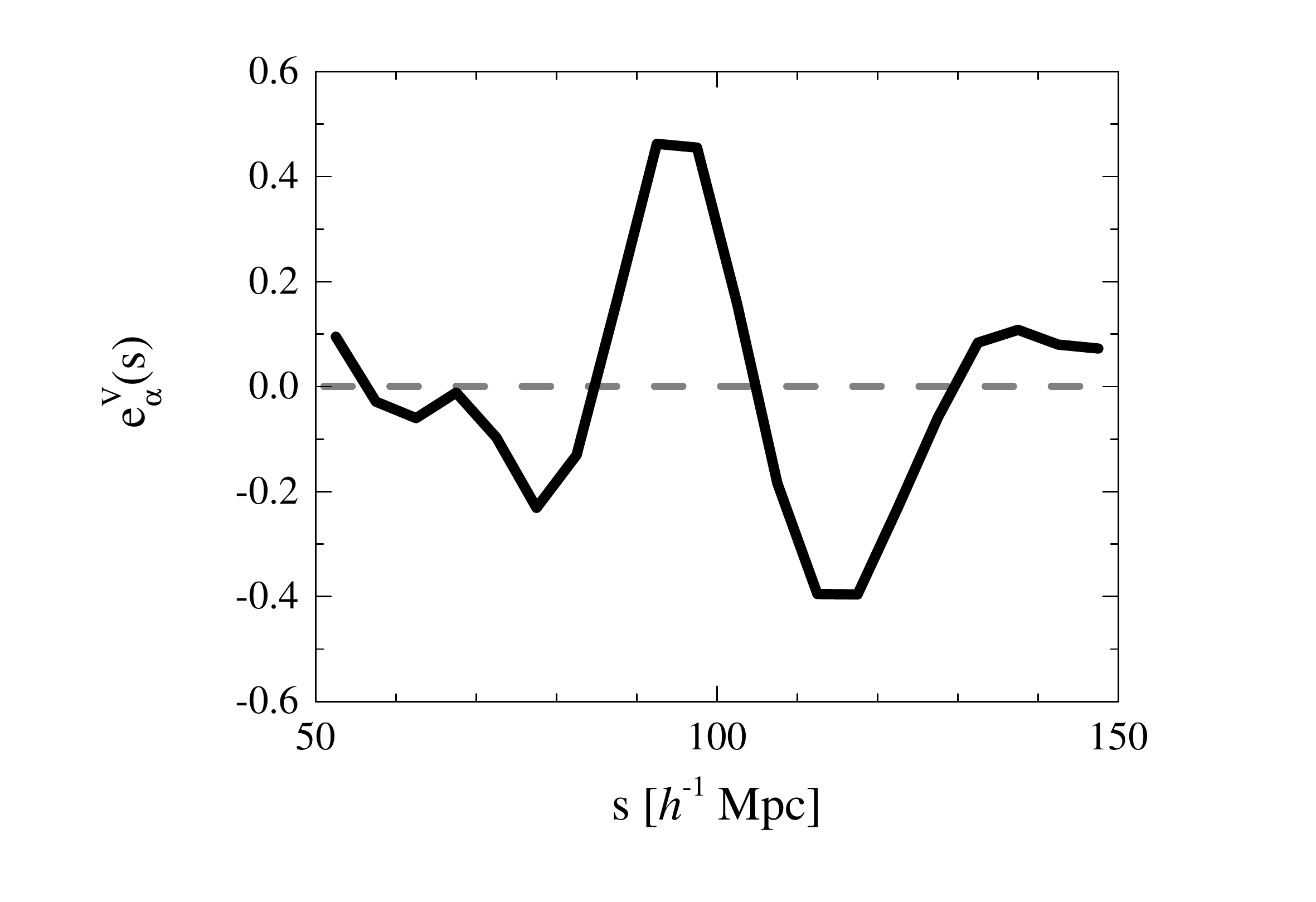}
\caption{\yt{ The projection vector for the $\alpha$ parameter ${\bf e}_\alpha^V$, as defined in Eq (\ref{eq:eV}).}}
\label{fig:fig4}
\end{figure} 

%\begin{figure*} %[!t]
%\centering
%\includegraphics[scale=0.18]{iso_FoM_Oct.pdf}
%\caption{Panel (A): The first four (with largest variances) and the last two (with least variances) eigenvectors of the data covariance matrix for the monopole of the galaxy correlation function; (B): The forecasted FoM of the isotropic BAO parameter $\alpha$, as a function of the number of total reordered PC modes used. The FoM is normalized so that it is exactly $1$ if all the PC modes are used; (C): A comparison between the measured $\xi_0$ (black curves with the 68\% CL error bars) and the reconstructed $\xi_0$ from the first ten PC modes (white curve with a blue band showing the 68\% CL uncertainty); (D): The projection vector for the $\alpha$ parameter ${\bf e}_\alpha^V$, as defined in Eq (\ref{eq:eV}).}
%\label{fig:iso_result}
%\end{figure*} 

\subsection{MOPED compression} 
\label{sec:moped}
The selected PC modes ($M'$) based on Fisher information matrix can further be compressed using the MOPED scheme, with almost no loss of information for the parameters we are interested in, and it can also highlight the key informative modes for the specific parameters we aim to constrain.   

We seek a transformation matrix ${\bf V}$ that compresses the $\gamma$'s without loss of information, which means that the compressed data vector ${\bm  \gamma}^{V}$ contains the same information as $\bm  \gamma$, \ie, 
\ba\label{eq:GammaV} 
{\bm  \gamma } ^V \equiv {\bf V}^T{\bm  \gamma }, \ {\bm  \gamma} \equiv\left[\gamma_1,\gamma_2,...,\gamma_{M'} \right]^T\,.
\ea 
Mathematically, this compression is lossless if the weighting vectors for multiple parameters $\{\theta_1, \theta_{p>1} \}$ satisfy \citep{MOPED},
\ba \label{eq:V}
{\bf v}_{1} &=& \frac{\bm{\Lambda}^{-1} {\bm \gamma}_{, 1}}{\sqrt{{\bm \gamma}_{, 1}^{T} \bm{\Lambda}^{-1} {\bm \gamma}_{, 1}}} \,, ~~{\bm \gamma}_{, 1} \equiv \frac{\partial \bm \gamma}{\partial\theta_1} \,,   \\
{\bf v}_{p>1} &=& \frac{\bm{\Lambda}^{-1}{\bm \gamma}_{, p}-\sum_{q=1}^{p-1}\left({\bm \gamma}_{, p}^{T} \mathbf{v}_{q}\right) \mathbf{v}_{q}}{\sqrt{{\bm \gamma}_{, p}^T \bm{\Lambda}^{-1} {\bm \gamma}_{, p}-\sum_{q=1}^{p-1}\left({\bm \gamma}_{, p}^{T} \mathbf{v}_{q}\right)^{2}}} \,, {\bm \gamma}_{, p} \equiv \frac{\partial \bm \gamma}{\partial\theta_p}  \nonumber
\ea
Note that the weighting vectors are orthonormalized as explained in \citep{MOPED}, such that the compressed $\gamma$'s are uncorrelated and have unit variances\footnote{One can also use the non-orthonormal weights via the Karhunen-Lo$\grave{e}$ve algorithm \citep{KL}, \ie\,${\bf v}_{q} = {\bm  \Lambda^{-1} \bm \gamma}_{,q} \,, {\bm \gamma}_{,q} \equiv \partial {\bm \gamma}/\partial\theta_q\,, q=1,...p$, which was used to compress the galaxy power spectrum and bispectrum \citep{Gualdi:2017iey}. In this case, a certain level of redundancy may exist, and the correlations between the compressed $\gamma$'s, are quantified by the covariance matrix ${\bf  C}_V  \equiv {\bf V}^T {\bm  \Lambda} {\bf V}$, here ${\bf V} \equiv \{{\bf v}_1, ..., {\bf v}_p\}$.}.

We can define the projection vector for a given parameter $p$ as 
\ba\label{eq:eV} 
{\bf e}_p^V \equiv \mathbf{v}^T_{p}{\bf W}'\,, 
\ea 
here, the matrix ${\bf W}'$ is the submatrix of ${\bf W}$ and stores the first $M'$ eigenvectors. Thus the compressed $\gamma_p$ is given by a scalar and can be written as   
\ba\label{eq:gammaV} 
\gamma_p^V= {\bf e}_p^V \cdot \xi({S})\,.
\ea 
The role of the projection vector ${\bf e}_p^V$ is to \yt{optimally} extract the relevant information content for parameter \yt{$\theta_p$} from $\xi(S)$ into one single number, \yt{as the shape of ${\bf e}_p^V$ directly shows the decisive feature about each parameter we seek for cosmological analysis in two-point CF.} It is straightforward to generalize to cases with multiple parameters, in which ${\bf e}^V \equiv \{{\bf e}_1^V, ..., {\bf e}_p^V\}$ has multiple columns (one for each parameter). 

\section{Demonstration} 
\label{sec:dem}
In this section, we present a demonstration of the method we developed, using the mock and actual galaxy catalogs from the BOSS DR12  \citep{BOSSDR12,patchy}. We apply the method on analyzing the isotropic CF and the anisotropic CF respectively.

\subsection{Isotropic BAO analysis} 
 We start from a measurement of the isotropic BAO distance scale in a redshift range of $z\in[0.40,0.55]$, using $\xi_0$, the monopole of the CF. A BAO analysis in this redshift slice has been performed using the same galaxy sample \citep{BOSS:2016zkm}, which can be used for a direct comparison. 

We measure $\xi_0$ from $2048$ realizations of the MutliDark-Patchy mock catalogs \citep{patchy}, from which we evaluate the data covariance matrix ${\bf C}$ using Eq.\,\ref{eq:cov}. \yt{The error induced by the estimation of covariance matrix from the finite number of mocks is accounted for via correction factors, \ie, Eq.\,(27-28) in \citet{BOSS:2016zkm} (for details see \citet{Percival:2013sga}).} We then perform a PCA of ${\bf C}$ using Eq.\,\ref{eq:PCA}, and show the first four and the last two eigenvectors (ordered by the variances) in \yt{Fig \ref{fig:fig1}}. As shown, the first few modes with the largest variances, which are expected to carry most of the signal in data, are rather smooth, while the last few noisy modes have the least variances.

We use the same template as in the DR12 analysis, \ie, 
\ba
\xi_0^{\rm th}(s) = B^2\xi_0(\alpha s)+{a_1}/{s^2}+{a_2}/{s}+a_3 \,,
\ea
where $\smash{\alpha= \left[D_V(z)r_{d}^{\rm fid} \right]/ \left[D^{\rm fid}_V(z)r_{d}\right]}$ is a dilation factor, describing an isotropic shift in the BAO scale. $D_V(z)$ is the isotropic volume distance, and $r_d$ is the sound horizon scale at the drag epoch. The superscripts ($\rm fid$) denote these two quantities in the fiducial cosmology, which is the same one as in \citet{BOSS:2016zkm}. $B$ denotes the bias factor, and $a_1,..., a_3$ are co-varied to marginalize over the broad-band shape. Instead of using $\xi_0$ as observables as in \citet{BOSS:2016zkm}, we use the $\gamma$'s, the coefficients of the PC modes. We use $\xi_0$ from $s=50$ to $150 \, h^{-1}\rm Mpc$ with the bin width of $\smash{\Delta s=5}\, h^{-1} \rm Mpc$, such that there are $20$ data points in the traditional analysis, or $20$ PC modes in total, as in \citep{BOSS:2016zkm}. To determine how many PC modes are sufficient to extract most of the information for \yt{the $\alpha$ parameter in $\xi_0$}, we first perform a forecast using the Fisher matrix, with free parameters including $\smash{ \mathbf{ \Theta}\equiv\{\alpha, B^2, a_{1-3}\}}$.

\begin{table} %[!t]
\centering
\begin{tabular}{c|cc}
\hline\hline
     & $\xi_0(s)$   &  $\gamma$'s       \\ \hline
Mocks &  $ 1.0033\pm0.0229 $    & \yt{$ 1.0029\pm0.0229 $}   \\\hline
Samples& $1.0234\pm0.0198$      &  \yt{$1.0261\pm0.0219$}    \\
\yt{$N_{\rm data}$} & \yt{$20$} & \yt{$8$} \\
\yt{$\chi^2_{\rm min}$} & \yt{$22.3$} & \yt{$4.2$} \\
\yt{$\chi^2_{\rm red.}$} & \yt{$1.5$} & \yt{$1.4$} \\
\hline\hline
\end{tabular}
\caption{The constraints on $\alpha$ from mocks and galaxy samples using $\xi_0$ in \citet{BOSS:2016zkm} and the PC modes in this work as observables, respectively. $N_{\rm data}$ is the number of data points used in two methods. \yt{$\chi^2_{\rm min}$ is the minimum chi$-$square. $\chi^2_{\rm red.} = \chi^2_{\rm min} / \nu$ is the reduced chi$-$square, where $\nu$ is the degrees of freedom given by the number of data points minus the number of fitted parameters.}}
\label{tab:isobao}
\end{table}

\begin{figure*} %[!t]
\centering
\includegraphics[scale=0.23]{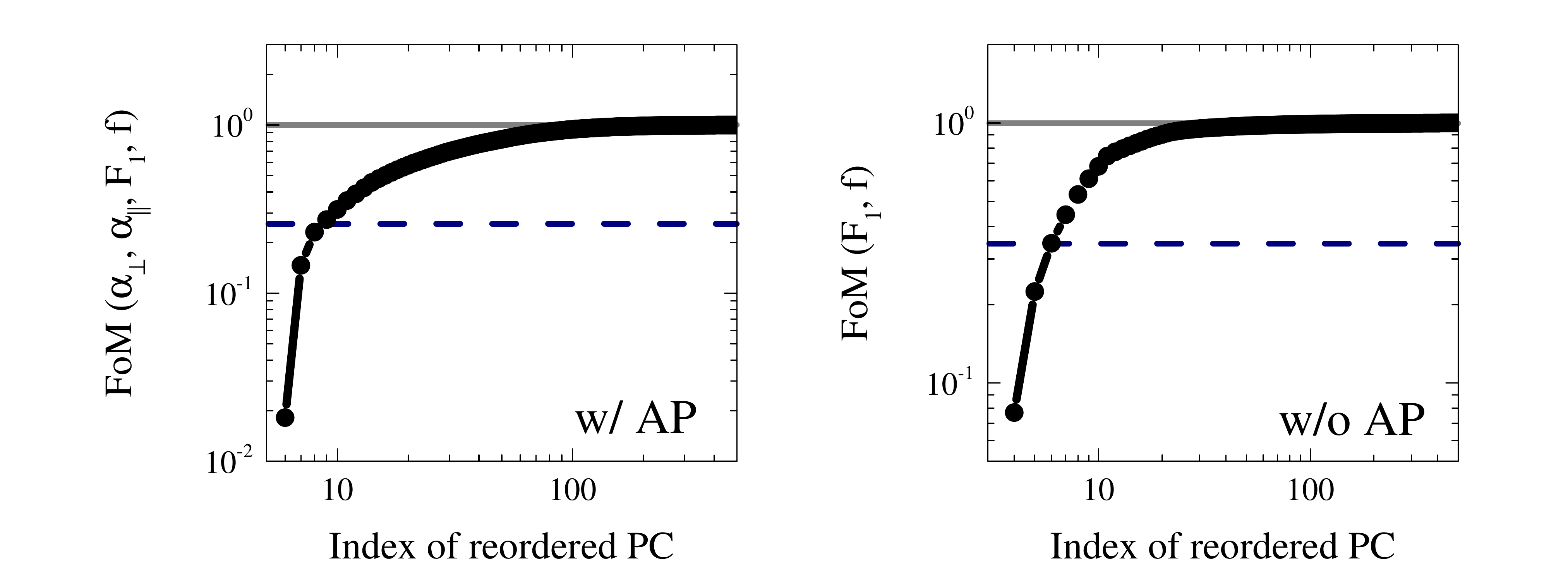}
\caption{\yt{Left panel: The FoM of $\{ \alpha_{\bot},\alpha_{||},F_1, f \}$ derived from various numbers of the total PC modes. The PC modes are re-ordered by the contribution of each mode to the FoM, and the FoM has been normalized so that it is unity using all the modes. The upper horizontal line shows the FoM using all the PC modes, and the lower dashed line shows the FoM using the monopole and quadrupole only; Right panel: same as left panel but with $\alpha_{\bot}=\alpha_{||}=1$.}}
\label{fig:fig5}
\end{figure*} 

\begin{figure*} %[!t]
\centering
\includegraphics[scale=0.23]{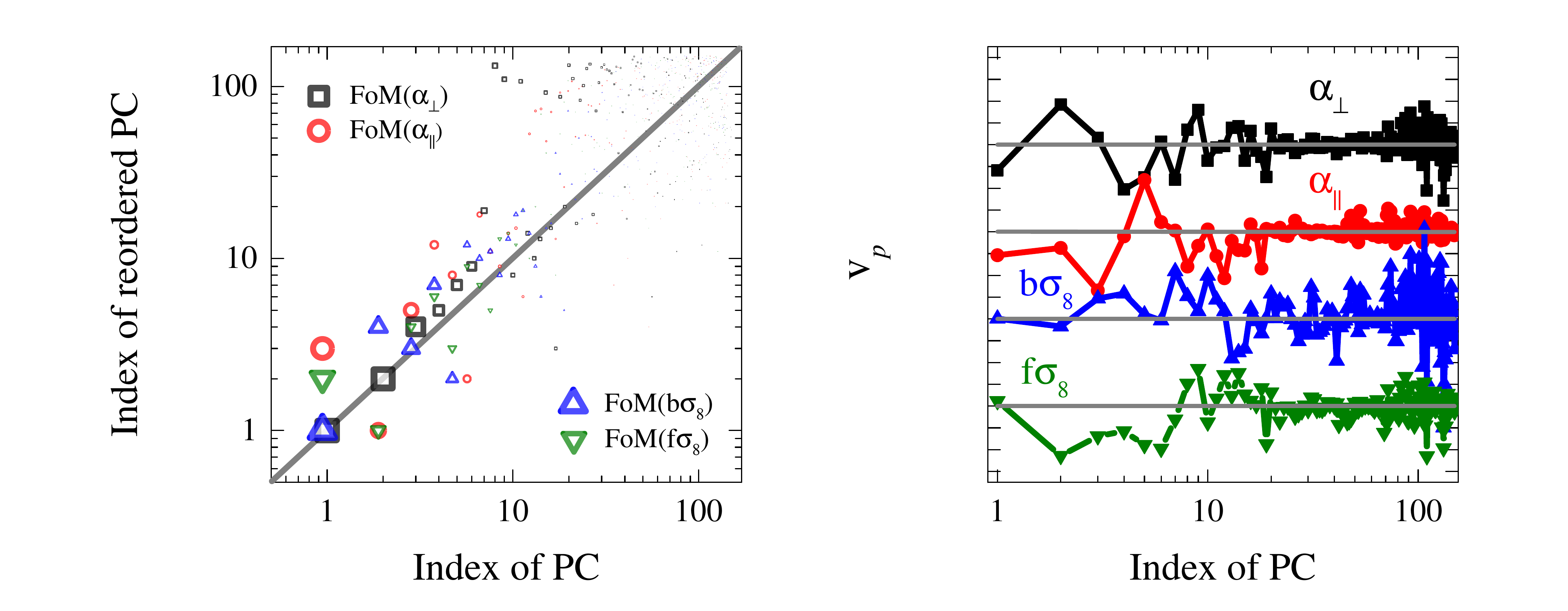}
\caption{\yt{Left panel: A scatter plot showing the correspondence between the re-ordered PC modes and the original PC modes. Different shapes represent the FoM for each parameter. For each parameter, the size of the symbols is proportional to the values of the FoM. The diagonal line shows the situation in which the modes ordered by the variance contribute to the FoM in the same order; Right panel: The optimal weights for various parameters, as shown in Eq. (\ref{eq:V}). The weights are arbitrarily offset for illustration, and the horizontal lines show ${\rm v}_p=0$.}}
\label{fig:fig6}
\end{figure*} 

We find that the original PC modes ordered by the variance of each mode, contribute to the Figure of Merit (FoM) defined as $1/\sqrt{\rm det\, Cov(\theta_1,\theta_2,...)}$ \citep{Albrecht:2009ct}, following almost the same order. In \yt{Fig \ref{fig:fig2}}, we show the forecasted FoM of $\alpha$, which is simply $1/\sigma(\alpha)$, with other parameters marginalized over, as a function of the re-ordered PC modes. It is evident that the FoM starts to saturate with \yt{$\smash{M'=8}$} modes, which is \yt{less than} half of the total number of data points in the traditional analysis. A direct comparison of the constraint on $\alpha$ using these two approaches is presented in Table \ref{tab:isobao}. As shown, the PCA approach returens almost \yt{the same BAO precision (\ie, $\sim2\%$)} to the traditional method, with the number of data points halved. This means that the last $12$ PC modes carry almost no information on $\alpha$, and they are also likely to be subject to observational systematics. \yt{The PCA method has a reduced chi$-$square of $1.4$, indicating a good fit to the first eight modes.} A reconstruction of $\xi_0(s)$ using the first \yt{eight} informative modes is shown in \yt{Fig \ref{fig:fig3}}, which is a slightly smoothed version of the original measurement, with the key features retained and uncertainty almost unchanged. 

To illustrate where the key information for $\alpha$ stems from, we plot the projection vector ${\bf e}_{\alpha}^V$, which is introduced in Eq.\,\ref{eq:eV}, in \yt{Fig \ref{fig:fig4}}. This vector effectively picks up information in $\xi_0$ in the range of $[80,130] \ h^{-1}\rm Mpc$ in a specific way, \ie, by contrasting $\xi_0$ around $105 \ h^{-1}\rm Mpc$, where the BAO peaks.

\subsection{Analysis of anisotropic correlation function}

In what follows, we consider the most general case, in which the anisotropic CF $\xi(s,\mu)$ is used as the raw observable, measured on a $\smash{N_s \times N_{\mu}}$ grid from the same BOSS DR12 catalogues, but in a wider redshift range ($\smash{0.5<z<0.75}$) to enhance the signal. Here, $\smash{N_s=(150-25)/5=25}$ and $\smash{N_{\mu}=(1-0)/0.05=20}$, denote numbers of the $s$ and $\mu$ bins, respectively. Given the number of data points and mocks available, we chose to use a semi-analytic approach to estimate the data covariance matrix in the presence of a non-uniform survey window function, with the tool of {\tt RascalC} \citep{Philcox:2019ued}. In this approach, the covariance matrix $\smash{C_{ij,k\ell}\equiv \langle \xi(s_i,\mu_j) \xi(s_k,\mu_{\ell})\rangle - \langle \xi(s_i,\mu_j) \rangle \langle \xi(s_k,\mu_{\ell})\rangle}$ is calculated using the galaxy and random catalogues as an input, with the non-Gaussian contribution approximated via a jackknife-rescaled inflation of shot-noise, which is shown to work well on cosmological analyses including the determination of the BAO scale. \citep{OConnell:2015src,OConnell:2018oqr,Philcox:2019ued}.

As we did for $\xi_0$, we perform a PCA on $C_{ij,k\ell}$ using Eq.\,\ref{eq:PCA}, and use the coefficients of the PC modes as observables to constrain BAO and RSD parameters in the set $\smash{\mathbf{ \Theta}\equiv\{\alpha_{\perp}, \alpha_{||}, F_1, f, F_2, \sigma_{\rm FoG}\}}$, where the scaling parameters $\smash{\alpha_{\perp}\equiv\left(D_A r_{d}^{\rm fid} \right)/ \left(D^{\rm fid}_A r_{d}\right)}$ and $\smash{\alpha_{||}\equiv \left(H^{\rm fid} r_{d}^{\rm fid} \right)/ \left(H r_{d}\right)}$ are introduced to account for the Alcock-Paczynski (AP) effect \citep{APtest}, $D^{\rm fid}_A$  and $H^{\rm fid}$ are the angular diameter distance and Hubble expansion rate in the fiducial cosmology, respectively. The quantities $F_1$ and $F_2$ are the local Lagrangian bias parameters (the Eulerian linear bias $b$ is related by $\smash{b=1+F_1}$), $f$ is the linear growth rate, and $\sigma_{\rm FoG}$ is used to marginalize over the Fingers-of-God (FoG) effect on nonlinear scales. The Gaussian Streaming model is used to link theory to observables \citep{Reid:2011ar, Reid2012}, 
\ba
1+\xi^{\rm th}(s_{\perp}, s_{\parallel}) &=& \int \frac{\d y }{\sqrt{2\pi \left[\sigma^2_{12}(r, \mu)+\sigma^2_{\rm FoG}\right]}}  \left[1+\xi(r)\right] \nonumber \\
&\times& \exp \left\{-\frac{\left[s_{\parallel} - y - \mu v_{12}(r)\right]^2}{2 \left[\sigma^2_{12}(r, \mu)+\sigma^2_{\rm FoG}\right] }\right\} \,,
\label{eq:streaming}
\ea
where $\smash{s_{||}\equiv s  \mu}$ and $\smash{s_{\perp}\equiv\sqrt{s  (1-\mu^2)}}$ denote the separation of pairs along and across the LOS, respectively, $\xi(r)$ is the real-space CF as a function of the real-space separation $r$, $v_{12}(r)$ is the mean infall velocity of galaxies separated by $r$, and $\sigma_{12}(r, \mu)$ is the pairwise velocity dispersion of galaxies. The quantities $\xi(r)$, $v_{12}(r)$ and $\sigma_{12}(r, \mu)$ are computed using the Convolution Lagrangian Perturbation Theory (CLPT) \citep{Carlson:2012bu, Wang:2013hwa}.

\begin{figure*} %[!t]
\centering
\includegraphics[scale=0.35]{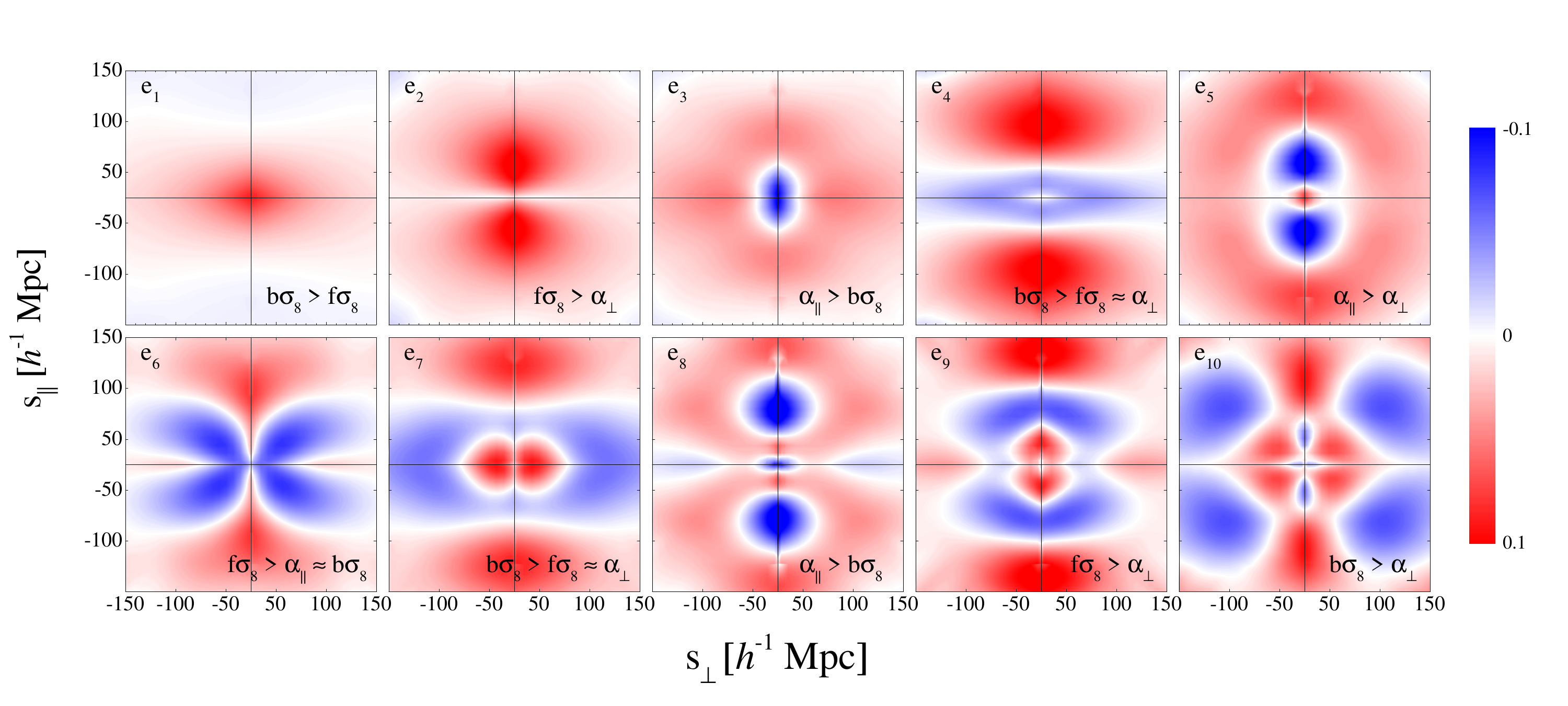}
\caption{The first ten eigenvectors of the data covariance matrix for the anisotropic CF $\xi(s,\mu)$. To quantify the information content in each mode for the cosmological parameters, we use each PC mode to constrain the four parameters (one at a time), and compute the efficiency of each mode to constrain each of the parameters. For each mode, we show the most relevant parameter, followed by the one(s) that are less relevant. See text for more details.}
\label{fig:aniso_evec}
\end{figure*}

\begin{figure*} %[!htb]
\centering
\includegraphics[scale=0.17]{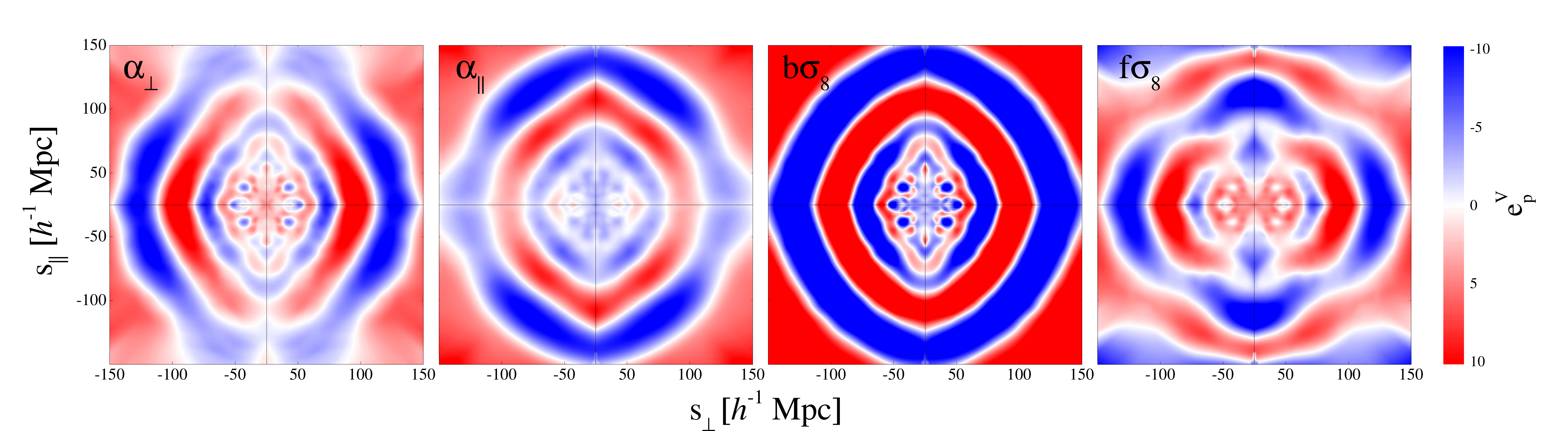}
\caption{The optimal combinations of the PC modes, ${\bf e}_p^V$ defined in Eq. (\ref{eq:gammaV}), that carry \yt{the decisive features for} the BAO and RSD parameters \yt{in two-point CF}, as illustrated in the legend.}
\label{fig:wevector}
\end{figure*}

\begin{figure} %[!b]
\centering
\includegraphics[scale=0.28]{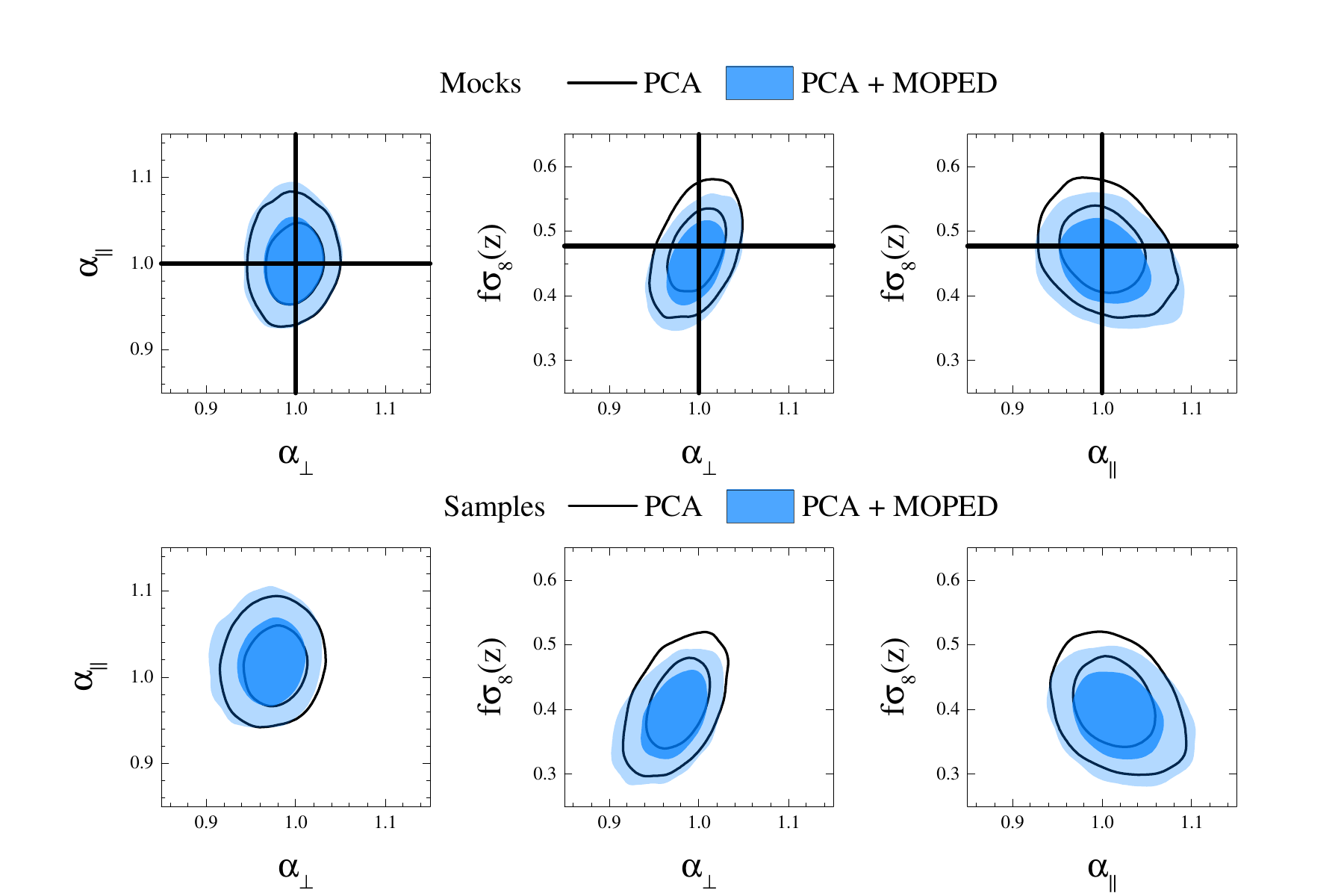}
\caption{The 68 and 95\% CL contours for the BAO and RSD parameters derived from the averaged mocks (upper panels) and from the BOSS DR12 galaxy sample (lower). Results using the original PC modes or the optimally combined modes as observables are shown with black unfilled contours and blue filled contours. Horizontal and vertical lines in the upper panels illustrate the fiducial model of the mocks, which is expected to be consistent with the contours.}
\label{fig:con}
\end{figure} 

A Fisher forecast is performed to determine the number of modes required, before the actual data analysis. \yt{Left panel of Fig \ref{fig:fig5}} shows the FoM for parameters $\{\alpha_{\perp}, \alpha_{||}, F_1, f\}$ as a function of numbers of the total PC modes used, and the modes are re-ordered by the contribution of each mode to the FoM. In \yt{left panel of Fig \ref{fig:fig6}} , we plot the indices of the PC modes before and after the re-ordering, and find that the modes ordered by the variance, \ie, the original PC modes, contribute to the FoM following a similar order, which confirms that the noisy PC modes with the smallest eigen-values are least informative for a BAO and/or RSD analysis. As illustrated, the FoM grows quickly with the number of modes (\eg, only $9$ modes are required to extract \yt{the same information as in $\xi_0$ and $\xi_2$}, which are measured in $50$ data points in the traditional method), and gets saturated with $\sim150$ modes (only $\sim25$ modes are needed if the AP effect is not considered, as shown in \yt{right panel of Fig \ref{fig:fig5}} .

The first ten eigenvectors ($e_1$ to $e_{10}$) are shown in Fig. \ref{fig:aniso_evec}, and interesting patterns show up in these modes: (a) all the modes are anisotropic, reflecting the fact that they all contain information for the AP and RSD effect; (b) there is some level of isotropy on large scale ($s\gtrsim100 \ h^{-1} \ {\rm Mpc}$) in some modes, \eg, $e_3$ and $e_5$, showing the impact of BAO; (c) the complexity of features increases with the index of modes, but a pattern exists, \eg, $e_1$ has no node (it does not cross zero), $e_2$ has one node in the $s_{||}$ direction, $e_3$ has one node in both $s_{||}$ and $s_{\bot}$ directions, \etc

To associate these modes with the BAO/RSD parameters we are interested in, we compute the constraint of each mode on each individual parameters using the Fisher matrix technique, and identify the most relevant parameter(s) for each mode, \ie, the parameter(s) that are extracted with the highest efficiency from each PC mode. In practice, we first compute the FoM of each parameter using all the modes to define ${\rm FoM}_{\rm tot}$, the total information content for each parameter, then compute the efficacy of using the $i$th mode by evaluating ${\rm FoM}_i / {\rm FoM}_{\rm tot}$. In Fig. \ref{fig:aniso_evec}, the most relevant parameters are shown for each mode, \eg, the apparently anisotropic modes $e_1,e_2,e_4,e_6,e_7,e_9$ and $e_{10}$ are crucial to determine the RSD parameters $f\sigma_8$ and $b\sigma_8$, while the quasi-isotropic modes $e_3,e_5$ and $e_8$ are more useful for $\alpha_{\bot}$ and $\alpha_{||}$.

As discussed previously, the PC modes can be optimally combined in a way such that \yt{the maximum information content for each parameter from PC modes} is stored in one single mode. We therefore compute the optimal weights using Eqs (\ref{eq:V})-(\ref{eq:gammaV}), and show the resultant modes for BAO and RSD parameters in Fig. \ref{fig:wevector}, with coefficients for the PC modes, ${\rm v}_p$, in \yt{right panel of Fig \ref{fig:fig6}} . We can tell that for all the parameters, the first ten modes contribute the most to $|{\rm v}_p|$, and $|{\rm v}_p|$ quickly decays roughly after the $20$th mode. The parameter-specific modes in Fig. \ref{fig:wevector} are rather intuitive: the BAO/AP modes, denoted as $\alpha_{\bot}$ and $\alpha_{||}$, have clear ring-like structures, which are quasi-isotropic around the BAO scale, with an anisotropy therein to reflect the role of each parameter, \ie, the $\alpha_{\bot}$ and $\alpha_{||}$ mode shows a pattern of upweights across and along the LOS, respectively, which is as expected. The modes for $f\sigma_8$ and $b\sigma_8$, however, carry a high level of anisotropy, \ie, an obvious squashing pattern along the LOS, highlighting the RSD effect. These modes \yt{contain the decisive features for the desired parameters, and can be used to optimally filter the information content for each parameter in two-point CF.} We also present the weighting eigenvectors with respect to another parametrization for BAO, \ie\, the isotropic shift $\alpha$ and the anisotropic warping factor $\epsilon$ in Fig. \ref{fig:MwevectorII} of the Appendix.

\begin{table} %[!t]
\centering
\footnotesize
\begin{tabular}{c|c|cc}
\hline\hline
&  $\xi_{\rm MQH} $   &   \multicolumn{2}{c}{$\xi(s,\mu) $}      \\
&                                & PCA  & PCA+MOPED\\\hline
Mocks &     &   & \\
$\Delta(\alpha_{\perp})$  & $-0.008      \pm  0.021$    &$-0.002	\pm	0.020$   & $-0.004 \pm 0.021$\\
$\Delta(\alpha_{\parallel})$  & $\,\,\,\,0.014    \pm 0.033$   &$\,\,\,\,0.002	\pm	0.031$ & $\,\,\,\,0.005 \pm 0.033$\\
$\Delta(f \sigma_8)$    & $-0.025     \pm  0.041 $  &$-0.006	\pm	0.042$  & $-0.025 \pm 0.042$ \\\hline
$\rm FoM$ &  $\,\,\,\,1.0$ &$\,\,\,\,1.0$& $\,\,\,\,0.9$\\
$\rm FoB$ &  $\,\,\,\,1.0$ &$\,\,\,\,0.22$ & $\,\,\,\,0.84$\\
\hline\hline
Samples&     &    & \\
 $\alpha_{\perp}$  & $\,\,\,\,0.961\pm0.024$    &$\,\,\,\,0.976\pm0.023$   & $\,\,\,\,0.971\pm0.025$ \\
$\alpha_{\parallel}$  & $\,\,\,\,1.034 \pm0.033$ &$\,\,\,\,1.015 \pm0.030$ & $\,\,\,\,1.020 \pm0.033$\\
$f\sigma_8$    &  $\,\,\,\,0.393 \pm 0.046$ &$\,\,\,\,0.408 \pm 0.045$  & $\,\,\,\,0.389 \pm 0.043$\\\hline
$\rm FoM$ &   $\,\,\,\,1.0$ &$\,\,\,\,\yt{1.17}$& $\,\,\,\,\yt{1.01}$ \\ 
\yt{$N_{\rm data}$} &\yt{$\,\,75$} & \yt{$\,\,\,\,150$}  & \yt{$6$} \\
\yt{$\chi^2_{\rm min}$} &\yt{$\,\,67$} & \yt{$\,\,\,\,144.6$}  & \yt{$\,\,\,\,4.8$} \\
\yt{$\chi^2_{\rm red.}$} & \yt{$\,\,0.97$} & \yt{$\,\,\,\,1.0$}  & \yt{$\,\,\,\,-$} \\
\hline\hline
\end{tabular}
\caption{Mean and 68\% CL uncertainty of the BAO and RSD parameters derived from the averaged mocks (upper part) and from the BOSS DR12 galaxy sample (lower). For the mock test, we show the deviation from the expected values. For each sample, results are shown using three kinds of observables: the monopole, quadrupole and hexadecapole ($\xi_{\rm MQH}$), the original PC modes derived from $\xi(s,\mu)$ (PCA), and the optimally combined PC modes (PCA+MOPED). The FoM and FoB are normalized using those derived from $\xi_{\rm MQH}$ so that ${\rm FoM} \ (\xi_{\rm MQH}) = {\rm FoB} \ (\xi_{\rm MQH}) =  1$. \yt{$N_{\rm data}$ is the number of data points used in each method. $\chi^2_{\rm min}$ is the minimum chi$-$square. $\chi^2_{\rm red.} = \chi^2_{\rm min} / \nu$ is the reduced chi$-$square, where $\nu$ is the degrees of freedom given by the number of data points minus the number of fitted parameters. }}
\label{tab:result}
\end{table}

To demonstrate our method, we apply our pipeline on the BOSS DR12 mock and galaxy samples described earlier, and present the main results in Table \ref{tab:result} and Fig. \ref{fig:con}. We constrain the BAO and RSD parameters using the original PC modes (up to $150$ PC modes, denoted as `PCA'), the optimally combined PC modes (denoted as `PCA+MOPED'), and the multipoles up to the hexadecapole (denoted as `$\xi_{\rm MQH}$') for a comparison, respectively, and find that the results are largely consistent with each other. As demonstrated by the averaged mocks, with the PCA approach, we are able to extract \yt{the same information as} in $\xi_0,\xi_2$ and $\xi_4$ quantified by the FoM using $150/500=30\%$ modes. In addition, the Figure of Bias (FoB) presented in Table \ref{tab:result}, which represents the significance of systematic biases relative to the statistical uncertainties, is defined as $\sqrt{\sum_{i,j}\Delta \theta_i {\rm F}_{ij}\Delta \theta_j}$ \citep{Shapiro:2008yk}, where $\Delta \theta_i$ is the deviation of the mean value from the expected one for the parameter $\theta_i$, and ${\rm F}_{ij}$ is the inverse of covariance matrix between parameters. It is found that FoB can be significantly reduced by a factor $4.5$ by the PCA method, largely due to the fact that the noisy modes are removed without loss of cosmological information. The PCA+MOPED approach, on the other hand, retains 90\% the FoM, probably due to the fact that the probability distribution of some parameters such as $\sigma_{\rm FoG}$ are non-Gaussian, making the compression defined in Eq. (\ref{eq:GammaV}) sub-optimal and thus subject to information loss \yt{or systematic bias} to some extent. But even in this situation, the FoB is reduced by 19\% \yt{compared with that by the $\xi_{\rm MQH}$ method}. 

When applied to the BOSS DR12 sample, we find that the FoM extracted by the PC modes is \yt{17\%} larger than that by the multipoles. \yt{Specifically, the FoM is computed via the covariance matrices from the $\xi_{\rm MQH}$ and PCA method for the $\{\alpha_{\perp}, \alpha_{||}, f\sigma_8\}$ parameters, \ie,
  \begin{equation}
\mathbf{C}_{\xi_{\rm MQH}} = 
 \begin{pmatrix}
5.61 \times 10^{-4} & 2.24 \times 10^{-5} & 4.47 \times 10^{-4} &  \\
 -  & 1.12 \times 10^{-3} & -5.48 \times 10^{-4} &  \\
 -  &  -  & 2.08 \times 10^{-3} &  \\
 \end{pmatrix} \,,
\end{equation}
and 
  \begin{equation}
\mathbf{C}_{\rm PCA} = 
 \begin{pmatrix}
5.35 \times 10^{-4} & 4.49 \times 10^{-5} & 4.63 \times 10^{-4} &  \\
 -  & 8.95 \times 10^{-4} & -4.11 \times 10^{-4} &  \\
 -  &  -  & 2.00 \times 10^{-3} &  \\
 \end{pmatrix} \,.
\end{equation}
Both $\xi_{\rm MQH}$ and PCA methods show a good fit to BOSS data. This is evidenced by the values of $\chi^2_{\rm red.}$ (close to unity).} The PCA+MOPED approach is also subject to information loss compared to the PCA method, but it extracts \yt{the same information as} in the multipoles. \yt{The corresponding covariance matrix from PCA+MOPED method is
 \begin{equation}
\mathbf{C}_{\rm PCA+MOPED} = 
 \begin{pmatrix}
4.5 \times 10^{-4} & 6.35 \times 10^{-5} & 3.88 \times 10^{-4} &  \\
 -  & 1.09 \times 10^{-3} & -3.85 \times 10^{-4} &  \\
 -  &  -  & 1.79 \times 10^{-3} &  \\
 \end{pmatrix} \,.
\end{equation}
Considering that the size of the compressed data vector in the PCA+MOPED approach is just the number of the fitted parameter itself, we only list the value of minimum chi$-$square in Table \ref{tab:result}.}

\section{Conclusion and Discussions} 
\label{sec:con}
In this era of precision cosmology, it is challenging to perform data analysis and cosmological implications for massive spectroscopic surveys, not only due to the large amount of observational data we are collecting, but also to the precision of measuring the cosmological parameters we require. This means that we need efficient and robust methods to extract cosmological information from galaxy surveys.

In this work, we develop a new method based on a principal component analysis for analyzing the anisotropic galaxy correlation functions $\xi(s,\mu)$. As we demonstrate using galaxy samples from BOSS DR12, our method is an ideal tool to meet our needs. A PCA of $\xi(s,\mu)$ can efficiently separate the cosmological signal we seek from the noise, which is non-informative. We confirm that most of the information for the BAO and RSD parameters, which are key for cosmological studies, is contained in the first few PC modes (the modes with the largest variances), making it natural to remove the PC modes with relatively small variances. Quantitatively, we find that using $150$ out of $500$ PC modes, one is able to \yt{extract key information in two-point CF} for a joint BAO and RSD study, and only $25$ out of $500$ modes are required for a RSD-only analysis. Being orthogonal to each other, the PC modes kept form a natural reservoir to store key information for general cosmological analysis, without redundancy. 

To associate the PC modes with the parameters of interest, we optimally combine the PC modes using the MOPED compression scheme, so that \yt{the maximum information in these PC modes} for each individual parameter is contained in one single mode. Thus the number of PC modes retained (\ie\,150 modes) can be reduced so that it is equal to the number of parameters of interest, \ie \, six parameters, with almost no loss of information. Our method and pipeline have been well validated using mock catalogs, and an application on the BOSS DR12 catalog shows that the FoM of BAO and RSD parameters is improved by $17\%$ compared to that using the traditional method.

Our method uses a mixture of simulations (for covariance matrix estimation) and analytic theory for the parameter dependence of the BAO and RSD effects. It therefore still depends to some extent on having a correct modeling for nonlinear effects. But the virtue of this framework is that it can be readily updated to incorporate revisions to theory. Our method provides an efficient way to extract the informative modes in the 2D correlation function and allows for a rapid estimation of BAO and RSD parameters from ongoing and forthcoming galaxy surveys including Dark Energy Spectroscopic Instrument (DESI) \citep{DESI}, the Euclid mission \citep{Euclid} and the Subaru Prime Focus Spectrograph (PFS) \citep{PFS}.

\section*{Acknowledgements}
We thank Ross O'Connell, Oliver H. E. Philcox and Daniel Eisenstein for help with the analytic covariance matrix. We also thank Oliver H. E. Philcox, Kwan Chuen Chan, Will Percival, Siddharth Satpathy, Martin White, Mike Wang and Shuo Yuan for helpful discussions. YW is supported by National Key R\&D Program of China No. (2022YFF0503404, 2023YFA1607800, 2023YFA1607803), NSFC grants (12273048, \yt{12422301}), the Youth Innovation Promotion Association CAS, and the Nebula Talents Program of NAOC. YW and GBZ are supported by the CAS Project for Young Scientists in Basic Research (No. YSBR-092), and NSFC Grants (1171001024, 11673025, 11890691). GBZ is supported by NSFC grant no. 11925303, science research grants from the China Manned Space Project with No. CMS-CSST-2021-B01, and the New Cornerstone Science Foundation through the XPLORER prize. JAP is supported by the European Research Council under grant no. 670193. This research used resources of the SCIAMA cluster funded by University of Portsmouth.

%%%%%%%%%%%%%%%%%%%%%%%%%%%%%%%%%%%%%%%%%%%%%%%%%%
\section*{Data Availability}
The correlation functions, covariance matrix, eigenvalues, eigenvectors  and weighting vectors are available at \url{https://github.com/ytcosmo/DataCompressionCF}.

\bibliographystyle{mnras}
\bibliography{main}

\begin{figure*} %[!htb]
\centering
\includegraphics[scale=0.17]{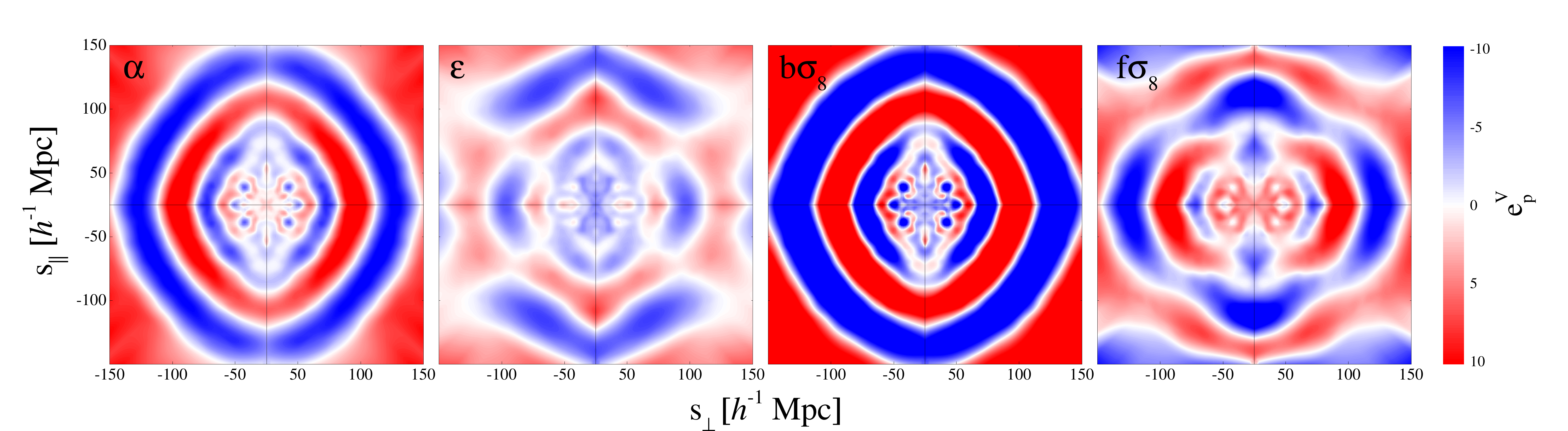}
\caption{Same as Fig. \ref{fig:wevector}, but for another BAO parametrization $\{\alpha, \epsilon$\}.}
\label{fig:MwevectorII}
\end{figure*}

\newpage
\appendix
Another parametrization for BAO is the isotropic shift $\alpha$ and the anisotropic warping factor $\epsilon$, which are defined in terms of $\alpha_{\perp}$ and $\alpha_{||}$, \ie\, $\alpha = \alpha_{\perp}^{2/3}\alpha_{||}^{1/3}$ and $\epsilon = (\alpha_{||}/\alpha_{\perp})^{1/3}-1$. The optimal combinations of PC modes for $\{\alpha, \epsilon, b\sigma_8, f\sigma_8\}$ are shown in Fig. \ref{fig:MwevectorII}. It is seen that the $\alpha$ mode manifests a clear isotropic pattern. In contrast, the $\epsilon$ mode has an obvious anisotropy. 

\bsp
\label{lastpage}

\end{document}